\documentclass[11pt,]{article}
\usepackage{lmodern}
\usepackage{amssymb,amsmath}
\usepackage{ifxetex,ifluatex}
\usepackage{fixltx2e} 
\ifnum 0\ifxetex 1\fi\ifluatex 1\fi=0 
  \usepackage[T1]{fontenc}
  \usepackage[utf8]{inputenc}
\else 
  \ifxetex
    \usepackage{mathspec}
  \else
    \usepackage{fontspec}
  \fi
  \defaultfontfeatures{Ligatures=TeX,Scale=MatchLowercase}
\fi
\IfFileExists{upquote.sty}{\usepackage{upquote}}{}
\IfFileExists{microtype.sty}{%
\usepackage{microtype}
\UseMicrotypeSet[protrusion]{basicmath} 
}{}
\usepackage[margin=1in]{geometry}
\usepackage{hyperref}
\hypersetup{unicode=true,
            pdftitle={Multilevel mixed effects parametric survival analysis: Estimation, simulation and application},
            pdfauthor={University of Leicester},
            pdfborder={0 0 0},
            breaklinks=true}
\usepackage{natbib}
\usepackage{graphicx,grffile}
\makeatletter
\def\maxwidth{\ifdim\Gin@nat@width>\linewidth\linewidth\else\Gin@nat@width\fi}
\def\maxheight{\ifdim\Gin@nat@height>\textheight\textheight\else\Gin@nat@height\fi}
\makeatother
\setkeys{Gin}{width=\maxwidth,height=\maxheight,keepaspectratio}
\IfFileExists{parskip.sty}{%
\usepackage{parskip}
}{
\setlength{\parindent}{0pt}
\setlength{\parskip}{6pt plus 2pt minus 1pt}
}
\ifx\paragraph\undefined\else
\let\oldparagraph\paragraph
\renewcommand{\paragraph}[1]{\oldparagraph{#1}\mbox{}}
\fi
\ifx\subparagraph\undefined\else
\let\oldsubparagraph\subparagraph
\renewcommand{\subparagraph}[1]{\oldsubparagraph{#1}\mbox{}}
\fi

\let\rmarkdownfootnote\footnote%
\def\footnote{\protect\rmarkdownfootnote}

\usepackage{titling}

\providecommand{\subtitle}[1]{
  \posttitle{
    \begin{center}\large#1\end{center}
    }
}

  \title{Multilevel mixed effects parametric survival analysis: Estimation,
simulation and application}
    \pretitle{\vspace{\droptitle}\centering\huge}
  \posttitle{\par}
  \subtitle{Michael J. Crowther}
  \author{University of Leicester}
    \preauthor{\centering\large\emph}
  \postauthor{\par}
      \predate{\centering\large\emph}
  \postdate{\par}
    \date{\href{mailto:michael.crowther@le.ac.uk}{\nolinkurl{michael.crowther@le.ac.uk}}}

\usepackage{stata}
\usepackage{bm}
\usepackage{xcolor}

\begin{document}
\maketitle
\begin{abstract}
In this article, I present the user written \texttt{stmixed} command for
the fitting of multilevel survival models, which serves as both an
alternative to Stata's official \texttt{mestreg}, and a complimentary
program with substantial extensions. \texttt{stmixed} can fit multilevel
survival models with any number of levels and random effects at each
level, including flexible spline-based approaches (such as
Royston-Parmar and the log hazard equivalent) or user-defined hazard
models. Simple or complex time-dependent effects can be included, as
well as the addition of expected mortality for a relative survival
model. Left-truncation/delayed entry can be used and \(t\)-distributed
random effects are provided as an alternative to Gaussian random
effects. The methods are illustrated with a commonly used dataset of
patients with kidney disease suffering recurrent infections, and a
simulated example, illustrating a simple approach to simulating
clustered survival data using \texttt{survsim}
\citep{Crowther2012b, CrowtherSurvsim}. \texttt{stmixed} is part of the
\texttt{merlin} family \citep{Crowther2017c, Crowther2018}.
\end{abstract}

\begin{center}
\today
\end{center}

\section{Introduction}\label{introduction}

Clustered survival data is often observed in a variety of settings.
Within medical research, a common example is the analysis of recurrent
event data, where individual patients can experience the event of
interest multiple times throughout the follow-up period, and the
inherent correlation within patients can be accounted for using a
frailty term \citep{Gutierrez2002}.

In the field of meta-analysis, the individual patient data (IPD)
meta-analysis of survival data is growing in use, as this form of
analysis is recognised as the gold standard approach
\citep{Simmonds2005}. Analysing the IPD simultaneously within a
hierarchical structure, allows direct adjustment for confounders and
incorporation of non-proportional hazards in covariate effects
\citep{Smith2005a, Crowther2012a, Crowther2014a}. Often a random
treatment effect is assumed to account for heterogeneity present in
treatment effects across the pooled trials.

A further area of interest is relative survival. Particularly prevalent
in cancer survival studies, relative survival allows the modelling of
excess mortality associated with a diseased population compared to that
of the general population \citep{Dickman2004}. Such data often exhibits
a hierarchical structure, with patients nested within geographical
regions such as counties. Patients living in the same area may share
unobserved characteristics, such as environmental aspects or medical
care access \citep{Charvat2016}.

With the release of Stata 14 came the \texttt{mestreg} command to fit
multilevel mixed effects parametric survival models, assuming normally
distributed random effects, estimated with maximum likelihood utilising
Gaussian quadrature. In this article, I present the user written
\texttt{stmixed} command for the fitting of multilevel survival models,
which serves as both an alternative to Stata's official
\texttt{mestreg}, and a complimentary program with substantial
extensions. \texttt{stmixed} can fit multilevel survival models with any
number of levels and random effects at each level, including flexible
spline-based approaches (such as Royston-Parmar and the log hazard
equivalent) or user-defined hazard models. Simple or complex
time-dependent effects can be included, as well as the addition of
expected mortality for a relative survival model.
Left-truncation/delayed entry can be used and \(t\)-distributed random
effects are provided as an alternative to Gaussian random effects.
\texttt{stmixed} has the ability to estimate a multilevel survival
model, with any of the aforementioned extensions, combined with a
user-defined hazard model, providing a platform for methods development
within a survival context.

In essence, \texttt{stmixed} is now a wrapper function for the recently
introduced \texttt{merlin} command \citep{Crowther2017c, Crowther2018},
which provides a general framework for estimating multivariate mixed
effects models. Multilevel survival models are one of the many classes
of models of which \texttt{merlin} can fit, but to make the methods more
accessible to researchers, I provide \texttt{stmixed} as a convenience
wrapper function, with a far simpler syntax, yet still providing the
power and flexibility of \texttt{merlin}.

The article is arranged as follows. Section \ref{sec:methods} describes
the multilevel parametric survival framework, and derives the likelihood
used to estimate the models, including the extension to relative
survival models and delayed entry/left-truncation. Section
\ref{sec:syntax} details the model syntax of \texttt{stmixed},
describing the available options, and Section \ref{sec:post} describes
the post-estimation tools available. I illustrate the command in Section
\ref{sec:eg} with a dataset of patients with kidney disease who are
followed up for recurrent infection at the catheter insertion point, and
show how to simulate clustered survival data using the \texttt{survsim}
command, representing an IPD meta-analysis scenario, with a random
treatment effect. I conclude the paper in Section \ref{sec:conc}.

\section{Multilevel mixed effects survival models}
\label{sec:methods}

For ease of exposition, I describe the methods in the context of a
two-level model, but \texttt{stmixed} can handle any number of levels. I
begin with some notation. Define \(i=1,\dots,N\) clusters (e.g.~trials
or centres), with each cluster having \(j=1,\dots,n_{i}\) patients. Let
\(S_{ij}\) be the true survival time of the \(j^{th}\) patient in the
\(i^{th}\) cluster, \(T_{ij}=\text{min}(S_{ij},C_{ij})\) the observed
survival time, with \(C_{ij}\) the censoring time. I define an event
indicator \(d_{ij}\), which takes the value of 1 if
\(S_{ij} \leq C_{ij}\) and 0 otherwise.

\subsection{Proportional hazards parametric survival models}
\label{sec:phmodels}

The proportional hazards mixed effect survival model can be written as
follows,

\begin{equation}
    h_{ij}(t) = h_{0}(t) \exp \left[ x_{ij}^{T}\beta + z_{ij}^{T}b_{j} \right]
    \label{eqn:hazph}
\end{equation}

where \(h_{0}(t)\) is the baseline hazard function of either a standard
parametric model, such as the exponential, Weibull or Gompertz
distributions, or a more general spline based approach, such as using
restricted cubic splines on the log hazard scale \citep{Bower2016a}, or
even a user-defined function. I define design matrices \(x_{ij}\) and
\(z_{ij}\) for the fixed (\(\beta\)) and random (\(b_{j}\)) effects,
respectively. I assume the random effects follow a multivariate normal
distribution, with \(b_{j} \sim \mbox{N}(0,\Sigma)\) (\texttt{stmixed}
also allowed multivariate \(t\)-distributed random effects). If
\(z_{ij} = 1\) for all \(i\) and \(j\), then Equation (\ref{eqn:hazph})
reduces to a shared frailty model, such as those available in
\texttt{streg}, albeit with a different choice of frailty distribution.

\subsection{Flexible parametric models}

An alternative to the standard proportional hazards distributions is the
flexible parametric model of \citet{Royston2002}, modelled on the
cumulative hazard scale, which has recently been extended to incorporate
random effects by \citet{Crowther2014a}. Therefore,

\begin{equation}
    H_{ij}(t) = H_{0}(t) \exp \left[ x_{ij}^{T} \beta + z_{ij}^{T}b_{j} \right]
    \label{eqn:surv}
\end{equation}

where \(H_{0}(t)\) is the cumulative baseline hazard function. The
spline basis for this specification is derived from the log cumulative
hazard function of a Weibull proportional hazards model. The linear
relationship with log time is relaxed through the use of restricted
cubic splines. Further details can be found in \citep{Royston2002} and
\citep{FPMbook}. On the log cumulative hazard scale, we have,

\begin{equation}
    \log \{ H_{ij}(t) \} = \eta_{ij}(t) = s\{\log(t)|\gamma,\textbf{k}_{0}\} + x_{ij}^{T} \beta + z_{ij}^{T}b_{j}
    \label{eqn:surv2}
\end{equation}

where \(s()\) are our basis functions, with knot vector
\(\textbf{k}_{0}\). Transforming to the hazard and survival scales,
gives

\begin{equation}
h_{ij}(t) = \left[\frac{1}{t}\frac{d s\{\log(t)|\gamma,\textbf{k}_{0}\}}{d \log(t)} \right]\exp(\eta_{ij}(t)), \qquad S_{ij}(t) = \exp [ -\exp(\eta_{ij}(t))]
    \label{eqn:surv3}
\end{equation}

In this framework I am assuming proportional cumulative hazards;
however, this in fact implies proportional hazards, as in the models
described in Section \ref{sec:phmodels}.

\subsubsection{Non-proportional (cumulative) hazards}

Relaxing the assumption of proportional hazards allows the investigation
of whether the effect of a covariate changes with time. Termed
non-proportional hazards, or time-dependent effects, the occurrence of
which is commonplace in the analysis of survival data. Examples include
treatment effects which vary over time \citep{Mok2009a}, and in registry
based studies, where follow-up can be substantial, covariate effects
have been found to vary \citep{Lambert2011}.

Non-proportional cumulative hazards have been incorporated into the
flexible parametric framework by \citet{Royston2002}, achieved by
interacting covariates with spline functions of log time and including
them in the linear predictor \citep{Lambert2009}. This provides even
greater flexibility in capturing complex effects, not restricted to
linear functions of time. Equation (\ref{eqn:surv2}) becomes

\begin{equation}
    \log \{ H_{ij}(t) \} = \eta_{ij}(t) = s\{\log(t)| \gamma,k_{0}\} +  x_{ij}^{T} \beta + z_{ij}^{T}b_{j}  + \sum_{r=1}^{R} s\{\log(t)|\delta_{r},k_{r}\} x_{ijr}
    \label{eqn:fpmnph}
\end{equation}

Each time-dependent effect can have varying number of spline terms,
depending on the number of knots, \(\textbf{k}_{r}\).

Within \texttt{stmixed}, time-dependent effects using restricted cubic
splines can be used with all available models.

\subsection{User-defined survival models}
\label{sec:user}

\texttt{stmixed} also allows the user to provide their own definitions
for the hazard function, with or without also defining a cumulative
hazard function, to allow the use of bespoke survival models with
general hazard functions. For estimation, both the hazard and cumulative
hazard functions are required (see Section \ref{sec:est} for further
details), so when only the hazard is provided, the cumulative hazard
function is calculated using numerical integration. Such a general
implementation allows the user complete flexibility, whilst still being
able to make use of the random effects engine within \texttt{stmixed},
which also syncs with the relative survival extension and delayed entry.
More details on writing a user-defined function can be found in
\citet{Crowther2018}.

\subsection{Relative survival}

Relative survival allows us to model the excess mortality associated
with a diseased population compared to that of the general population,
matched appropriately on the main factors associated with patient
survival, such as age and gender \citep{Dickman2004}. For a recent
extensive description and implementation of the tools available for
relative survival analysis in Stata, along with a description of the
differing approaches, I refer the reader to \citet{Dickman2015} and
references therein.

Concentrating on applications of relative survival to cancer settings,
the data generally comes from population based registries. Such data
often exhibits a hierarchical structure, with patients nested within
geographical regions such as counties. Patients living in the same area
may share unobserved characteristics, such as environmental aspects or
medical care access. \citet{Charvat2016} recently described a flexible
relative survival model allowing a random intercept, with the baseline
log hazard function modelled with B-splines, or restricted cubic
splines. In this article, I extend the multilevel Royston-Parmar
survival model described in \citet{Crowther2014a}, and essentially any
other hazard based survival model, to the relative survival setting,
further allowing any number of random effects, including random
coefficients. Modelling on the log cumulative hazard scale avoids the
need for numerical integration which is required when modeling on the
log hazard scale with splines, and will generally require fewer spline
terms than when modelling on the log hazard scale.

Within a multilevel modeling framework, I therefore define the total
hazard at the time since diagnosis, \(t\), for the \(j^{th}\) patient in
the \(i^{th}\) cluster (area) to be \(h_{ij}(t)\), with \[
    h_{ij}(t) = h_{ij}^{*}(t) + \lambda_{ij}(t)
\] where

\begin{itemize}
    \item $h_{ij}^{*}(t)$ is the expected mortality for the $j^{th}$ patient in the $i^{th}$ cluster (area)
    \item $\lambda_{ij}(t)$ is the excess mortality for the $j^{th}$ patient in the $i^{th}$ cluster (area)
\end{itemize}

and our model is \[
    \lambda_{ij}(t) = \lambda_{0}(t) \exp (X_{ij}^{T} \beta + Z_{ij}^{T}b_{i})
\] where \(\lambda_{0}(t)\) is the baseline hazard function, with
choices available including the exponential, Weibull, Gompertz,
spline-based or user-defined.

Alternatively, I could model on the (log) cumulative excess hazard
scale, using the flexible parametric model of \citet{Royston2002}, where
I define the total cumulative hazard at the time since diagnosis, \(t\),
for the \(j^{th}\) patient in the \(i^{th}\) cluster (area) to be
\(h_{ij}(t)\), with \[
    H_{ij}(t) = H_{ij}^{*}(t) + \Lambda_{ij}(t)
\] where

\begin{itemize}
    \item $H_{ij}^{*}(t)$ is the expected cumulative mortality for the $j^{th}$ patient in the $i^{th}$ cluster (area)
    \item $\Lambda_{ij}(t)$ is the excess cumulative mortality for the $j^{th}$ patient in the $i^{th}$ cluster (area)
\end{itemize}

and our model is \[
    \Lambda_{ij}(t) = \Lambda_{0}(t) \exp (X_{ij}^{T} \beta + Z_{ij}^{T}b_{i})
\] where \(\Lambda_{0}(t)\) is the baseline cumulative hazard function,
modeled with restricted cubic splines.

\subsection{Likelihood and estimation}
\label{sec:est}

Defining the likelihood for the \(i^{th}\) cluster under the mixed
effects framework, I define

\begin{equation}
        L_{i}(\theta) = \displaystyle \int_{-\infty}^{\infty} \left[ \prod^{n_{i}}_{j=1} p(T_{ij},d_{ij}| b_{i},\theta)\right] p(b_{i} | \theta) \hspace{0.75ex} \mbox{d} b_{i}
        \label{eqn:jl}
    \end{equation}

with parameter vector \(\theta\). Under a hazard scale model

\begin{equation}
        p(T_{ij},d_{ij}|b_{i},\theta) = h(T_{ij})^{d_{ij}} \exp \left [ - \int_{0}^{T_{ij}} h(u) \text{ d}u\right]
        \label{eqn:hazlike}
    \end{equation}

with \(h(T_{ij})\) defined in Equation (\ref{eqn:hazph}). Under the
flexible parametric survival model

\begin{equation}
        p(T_{ij},d_{ij}|b_{i},\theta) = \left [ \left \{\frac{1}{T_{ij}}\frac{\mbox{d}s\{\log(T_{ij})|\gamma,k_{0}\}}{\mbox{d} \log(T_{ij})}  \right \} \exp(\eta_{ij}) \right ]^{d_{ij}}\exp   \left \{-\exp(\eta_{ij}) \right \}
        \label{eqn:cumhazlike}
    \end{equation}

Finally, I assume the random effects follow a multivariate normal
distribution

\begin{equation}
    p(b_{i}|\theta) = (2\pi |\Sigma|)^{-q/2} \exp \left \{  -\frac{b_{j}^{\prime}\Sigma^{-1}b_{j}}{2}  \right \},
\end{equation}

with variance-covariance matrix, \(\Sigma\), and \(q\) the number of
random effects. The (possibly multi-dimensional) integral in Equation
(\ref{eqn:jl}) is analytically intractable, requiring numerical
techniques to evaluate. \texttt{stmixed} uses either \(m\)-point
mean-variance adaptive or non-adaptive Gauss-Hermite quadrature
\citep{Pinheiro1995, RHgllamm, Liu2008c}, or Monte-Carlo integration.
The default estimation method first fits the appropriate fixed effects
model, followed by the full model with variance and covariance
parameters given starting values of 0 and 1, respectively.
\texttt{stmixed} also allows multivariate \(t\)-distributed random
effects, with specified degrees of freedom, in which case only
Monte-Carlo integration is supported.

\subsection{Relative survival likelihood}

The adaption to the likelihood in Equation (\ref{eqn:jl}) to turn it
into a relative survival model is relatively simple. All that is needed
is the expected mortality rate at each event time, which are usually
obtained from national/regional life tables. Under a hazard scale model,
Equation (\ref{eqn:hazlike}) becomes,

\begin{equation}
        p(T_{ij},d_{ij}|b_{i},\theta) = \left [ h^{*}(T_{ij}) + \lambda(T_{ij}) \right ]^{d_{ij}} \exp \left [ - \int_{0}^{T_{ij}} \lambda(u) \text{ d}u \right]
        \label{eqn:hazlikerel}
    \end{equation}

and under a cumulative hazard scale model, Equation
(\ref{eqn:cumhazlike}) becomes,

\begin{equation}
        p(T_{ij},d_{ij}|b_{i},\theta) = \left [ \left \{\frac{1}{T_{ij}}\frac{\mbox{d}s\{\log(T_{ij})|\gamma,k_{0}\}}{\mbox{d} \log(T_{ij})}  \right \} \exp(\eta_{ij}) \right ]^{d_{ij}}\exp   \left \{-\exp(\eta_{ij}) \right \}
        \label{eqn:cumhazlike}
    \end{equation}

which provides substantial extensions to the relative survival
literature.

\subsection{Left-truncation/delayed entry}

The addition of left-truncation/delayed entry within a random effects
survival setting raises a particular extra level of complexity. The
random effects distributions are defined at \(t=0\), and as such, the
left truncation time point is conditional on on each patient's
subject-specific random effect contributions. For more details I refer
the reader to \citet{vandenBerg}. As such, our likelihood function
becomes,

\begin{equation}
        L_{i}(\theta | T_{0i}) = \frac{\displaystyle \int_{-\infty}^{\infty} \left[ \prod^{n_{i}}_{j=1} p(T_{ij},d_{ij}| b_{i},\theta)\right] p(b_{i} | \theta) \hspace{0.75ex} \mbox{d} b_{i}}{S(T_{0i} | \theta)}
        \label{eqn:lde}
\end{equation}

where \(S(T_{0i} | \theta)\) is the marginal survival function at the
entry time \(T_{0i}\), defined as

\begin{equation}
  S(T_{0i} | \theta) = \int_{-\infty}^{\infty} S(T_{0i} | b_{i}, \theta)p(b_{i} | \theta)  \mbox{ d} b_{i} \nonumber
\end{equation}

as such, there are two sets of analytically intractable integrals to
evaluate in Equation (\ref{eqn:lde}), which increases computation time.

\section{\texorpdfstring{The \texttt{stmixed}
command}{The stmixed command}}\label{the-stmixed-command}

\label{sec:syntax}

\subsection{Syntax}

\begin{stsyntax}
\hspace{2ex} stmixed
    \optional{\it{fe\_equation}}
    ||
    \it{re\_equation}
    \optional{|| \it{re\_equation}}
    \optional{,
    \texttt{\textit{options}}
}
\end{stsyntax}

where the syntax of \emph{fe\_equation} is

\begin{stsyntax}
\hspace{5ex}\optvarlist
\optif
\optin
\end{stsyntax}

and the syntax of \emph{re\_equation} is

\begin{stsyntax}
\hspace{5ex}\it{levelvar}: \optvarlist
\optional{,
\underbar{noc}onstant
}
\end{stsyntax}

\texttt{stmixed} requires that your data is \texttt{stset}.

\subsection{Options}

\subsubsection{Model}

\hangpara
\texttt{noconstant} suppresses the constant (intercept) term and may be
specified for the fixed effects equation and for the random effects
equations.

\hangpara
\texttt{distribution(string)} specifies the survival distribution.

\texttt{distribution(exponential)} fits an exponential survival model.

\texttt{distribution(weibull)} fits a Weibull survival model.

\texttt{distribution(gompertz)} fits a Gompertz survival model.

\texttt{distribution(rp)} fits a Royston-Parmar survival model. This is
a highly flexible fully parametric alternative to the Cox model,
modelled on the log cumulative hazard scale using restricted cubic
splines.

\texttt{distribution(rcs)} fits a log hazard scale flexible parametric
survival model. This is a highly flexible fully parametric alternative
to the Cox model, modelled on the log hazard scale using restricted
cubic splines.

\texttt{distribution(user)} specify a user-defined survival model; see
options below and \texttt{help\ merlin\ user-defined\ functions}.

\hangpara
\texttt{df(\#)} specifies the degrees of freedom for the restricted
cubic spline function used for the baseline function under a \texttt{rp}
or \texttt{rcs} survival model. \texttt{\#} must be between 1 and 10,
but usually a value between 1 and 5 is sufficient.The \texttt{knots()}
option is not applicable if the \texttt{df()} option is specified. The
knots are placed at the evenly spaced centiles ofthe distribution of the
uncensored log survival times. Note that these are
\texttt{interior\ knots} and there are also boundary knots placed at the
minimum and maximum of the distribution of uncensored survival times.

\hangpara
\texttt{knots(numlist)} specifies knot locations for the baseline
distribution function under a \texttt{rp} or \texttt{rcs} survival
model, as opposed to the default locations set by \texttt{df()}. Note
that the locations of the knots are placed on the standard time scale.
However, the scale used by the restricted cubic spline function is
always log time. Default knot positions are determined by the
\texttt{df()} option.

\hangpara
\texttt{tvc(varlist)} gives the name of the variables that have
time-varying coefficients. Time-dependent effects are fitted using
restricted cubic splines. The degrees of freedom are specified using the
\texttt{dftvc()} option.

\hangpara
\texttt{dftvc(df\_list)} gives the degrees of freedom for time-dependent
effects in \texttt{df\_list}. The potential degrees of freedom are
listed under the \texttt{df()} option. With 1 degree of freedom a linear
effect of log time is fitted. If there is more than one time-dependent
effect and different degress of freedom are requested for each
time-dependent effect then the following syntax applies:

\hangpara
\texttt{knotstvc(knotslist)} defines numlist \texttt{knotslist} as the
location of the interior knots for time-dependent effects.

\hangpara
\texttt{bhazard(varname)} specifies the variable which contains the
expected mortality rate, which invokes a relative survival model.

\hangpara
\texttt{covariance(vartype\_list)}, where each \texttt{vartype} is

\hangpara
\hspace{2ex} \texttt{diagonal} \texttt{\textbar{}} \texttt{exchangeable}
\texttt{\textbar{}} \texttt{identity} \texttt{\textbar{}}
\texttt{unstructured}

\hangpara
specifies the structure of the covariance matrix for the random effects.
An \texttt{diagonal} covariance structure allows a distinct variance for
each random effect within a random-effects equation and assumes that all
covariances are zero. \texttt{exchangeable} covariances have common
variances and one common pairwise covariance. \texttt{identity} is short
for ``multiple of the identity''; that is, all variances are equal and
all covariances are zero. \texttt{unstructured} allows for all variances
and covariances to be distinct. If an equation consists of \(p\)
random-effects terms, the \texttt{unstructured} covariance matrix will
have \(p(p+1)/2\) unique parameters. \texttt{covariance(diagonal)} is
the default.

\subsubsection{Integration}

\hangpara
\texttt{intmethod(intmethod)}, \texttt{intpoints(\#)}, and
\texttt{adaptopts(adaptopts)} affect how integration for the latent
variables is numerically calculated.

\hangpara
\texttt{intmethod(intmethod)} specifies the method and defaults to
\texttt{intmethod(mvaghermite)}. The current implementation uses
mean-variance adaptive quadrature at the highest level, and non-adaptive
at lower levels. Sometimes it is useful to fall back on the less
computationally intensive and less accurate \texttt{intmethod(ghermite)}
and then perhaps use one of the other more accurate methods.

\hangpara
\texttt{intmethod(mcarlo)} tells \texttt{stmixed} to use Monte-Carlo
integration, which either uses Halton sequences with
normally-distributed random effects, or anti-thetic random draws with
\(t\)-distributed random effects.

\hangpara
\texttt{intpoints(\#)} specifies the number of integration points to use
and defaults to \texttt{intpoints(7)} with
\texttt{intmethod(mvaghermite)} or \texttt{intmethod(ghermite)}, and
\texttt{intpoints(150)} with \texttt{intmethod(mcarlo)}. Increasing the
number increases accuracy but also increases computational time.
Computational time is roughly proportional to the number specified.

\hangpara
\texttt{adaptopts(adaptopts)} affects the adaptive part of adaptive
quadrature (another term for numerical integration) and thus is relevant
only for \texttt{intmethod(mvaghermite)}.

\hangpara
\texttt{adaptopts()} defaults to
\texttt{adaptopts(nolog\ iterate(1001)\ tolerance(1e-8))}.

\hangpara
\texttt{{[}no{]}log} specifies whether iteration logs are shown each
time a numerical integral is calculated.

\subsubsection{Estimation}

\hangpara
\texttt{from(matname)} allows you to specify starting values.

\hangpara
\texttt{restartvalues(sv\_list)} allows you to specify starting values
for specific random effect variances. See \texttt{merlin\ estimation}
for further details.

\hangpara
\texttt{apstartvalues(\#)} allows you to specify a starting value for
all ancillary parameters, i.e those defined by using the \texttt{nap()}
option.

\hangpara
\texttt{zeros} tells \texttt{stmixed} to use \texttt{0} for all
parameters starting values, rather than fit the fixed effect model. Both
\texttt{restartvalues()} and \texttt{apstartvalues()} can be used with
\texttt{zeros}.

\hangpara
\textit{maximization\_options}: \texttt{\underbar{dif}ficult},
\texttt{\underbar{tech}nique(algorithm\_spec)},
\texttt{\underbar{iter}ate(\num)},
{[}\texttt{\underbar{no}}{]}\texttt{\underbar{lo}g},
\texttt{\underbar{tr}ace}, \texttt{\underbar{grad}ient},
\texttt{showstep}, \texttt{\underbar{hess}ian},
\texttt{\underbar{shownr}tolerance},
\texttt{\underbar{tol}erance(\num)},
\texttt{\underbar{ltol}erance(\num)},
\texttt{\underbar{gtol}erance(\num)},
\texttt{\underbar{nrtol}erance(\num)},
\texttt{\underbar{nonrtol}erance}, \texttt{from(init\_specs)}; see
\rref{maximize}. These options are seldom used, but the
\texttt{difficult} option may be useful if there are convergence
problems.

\subsubsection{Reporting}

\hangpara
\texttt{showmerlin} displays the \texttt{merlin} syntax used to fit the
model.

\hangpara
\texttt{level(\#)} specifies the confidence level, as a percentage, for
confidence intervals. The default is \texttt{level(95)} or as set by
\texttt{set\ level}.

\section{\texorpdfstring{\texttt{stmixed}
postestimation}{stmixed postestimation}}\label{stmixed-postestimation}

\label{sec:post}

\subsection{Syntax for obtaining predictions}\begin{stsyntax}
predict
    \newvarname\
    \optif\
    \optin\
    \optional{,
    eta
    \underbar{h}azard
    \underbar{s}urvival
    \underbar{ch}azard
    cif
    rmst
    timelost
    fixedonly
    marginal
    at(\varname \hspace{0.5ex} \num \hspace{0.5ex} [\varname \hspace{0.5ex} \num \hspace{0.5ex}...])
    ci
    \underbar{time}var(\varname)
    level(\num)    
}
\end{stsyntax}

\subsection{Options}

\subsubsection{Predictions}

\hangpara
\texttt{eta} calculates the expected value of the linear predictor

\hangpara
\texttt{hazard} calculates the predicted hazard.

\hangpara
\texttt{survival} calculates each observation's predicted survival
probability.

\hangpara
\texttt{chazard} calculates the predicted cumulative hazard.

\hangpara
\texttt{cif} calculates the predicted cumulative incidence function.

\hangpara
\texttt{rmst} calculates the restricted mean survival time, i.e.~the
integral of the survival function up to time \(t\).

\hangpara
\texttt{timelost} calculates the time lost due to the event occuring,
i.e.~the integral of the cumulative incidence function up to time \(t\).

\subsubsection{Subsidiary}

\hangpara
\texttt{fixedonly} specifies predictions based on the fixed portion of
the model.

\hangpara
\texttt{marginal} specifies predictions calculated marginally with
respect to the random effects, i.e.~population-averaged predictions.

\hangpara
\texttt{at(\textit{varname} \num [\textit{varname} \num ...])} requests
that the covariates specified by the listed varname(s) be set to the
listed \# values. For example, \texttt{at(x1 1 x3 50)} would evaluate
predictions at x1 = 1 and x3 = 50. This is a useful way to obtain out of
sample predictions. Note that if \texttt{at()} is used together with
zeros all covariates not listed in \texttt{at()} are set to zero. If
\texttt{at()} is used without zeros then all covariates not listed in
\texttt{at()} are set to their sample values. See also \texttt{zeros}.

\hangpara
\texttt{ci} calculate confidence interval and store in
\textit{newvarname\_lci} and \textit{newvarname\_uci}. The delta-method
is used in all calculations using \texttt{predictnl}.

\hangpara
\texttt{timevar(\varlist)} defines the variable used as time in the
predictions. Default is \texttt{\_t}.

\hangpara
\texttt{level(\num)} specifies the confidence level, as a percentage,
for confidence intervals. The default is as set by \texttt{set level}.

\section{Examples}
\label{sec:eg}

In this section, I illustrate the command in two areas of research,
namely, recurrent events analysis, and the individual participant data
meta-analysis of survival data.

\subsection{Recurrent event data}

I consider the commonly used \texttt{catheter} dataset consisting of 38
patients with kidney disease \citep{McGilchrist1991}. The outcome of
interest is infection at the catheter insertion point, with our baseline
being time of initial catheter insertion. Patients can experience up to
two recurrences of infection, resulting in a total of 58 events. In the
examples I use the Royston-Parmar model for illustration. I intially fit
a null model, i.e.~no covariates and no random effects, to select the
degrees of freedom for the baseline cumulative hazard function, using
the Akaike Information Criterion to guide the choice. This selected 3
degrees of freedom (not shown), clearly indicating the need for a
flexible spline based model to capture the complex hazard function. I
now fit a Royston-Parmar proportional hazards model with a normally
distributed frailty, adjusting for age and gender,

\begin{verbatim}
. webuse catheter, clear
(Kidney data, McGilchrist and Aisbett, Biometrics, 1991)

. stset time, fail(infect)

     failure event:  infect != 0 & infect < .
obs. time interval:  (0, time]
 exit on or before:  failure

------------------------------------------------------------------------------
         76  total observations
          0  exclusions
------------------------------------------------------------------------------
         76  observations remaining, representing
         58  failures in single-record/single-failure data
      7,424  total analysis time at risk and under observation
                                                at risk from t =         0
                                     earliest observed entry t =         0
                                          last observed exit t =       562

. stmixed age female || patient:, distribution(rp) df(3)
Random effect M1: Intercept at level patient
variables created: _rcs1_1 to _rcs1_3

Fitting fixed effects model:

Fitting full model:

Iteration 0:   log likelihood = -332.01103  
Iteration 1:   log likelihood = -326.66495  
Iteration 2:   log likelihood = -326.12507  
Iteration 3:   log likelihood = -326.05683  
Iteration 4:   log likelihood = -326.05663  
Iteration 5:   log likelihood = -326.05663  

Mixed effects survival model                    Number of obs     =         76
Log likelihood = -326.05663
------------------------------------------------------------------------------
             |      Coef.   Std. Err.      z    P>|z|     [95% Conf. Interval]
-------------+----------------------------------------------------------------
_t:          |            
         age |   .0071588   .0129147     0.55   0.579    -.0181536    .0324712
      female |  -1.467426   .4928408    -2.98   0.003    -2.433376   -.5014758
 M1[patient] |          1          .        .       .            .           .
       _cons |  -.3469468   .6710705    -0.52   0.605    -1.662221    .9683271
-------------+----------------------------------------------------------------
patient:     |            
      sd(M1) |    .801154   .2732037                       .410625    1.563099
------------------------------------------------------------------------------
    Warning: Baseline spline coefficients not shown - use ml display
\end{verbatim}

Random effects are named with a \texttt{M} and a number. Therefore,
\texttt{stmixed} first provides some text ensuring the user can
understand which random effects correspond to what. It also reports
creating some spline variables, named \texttt{\_rcs\#\_1}, which are the
baseline splines for the Royston-Parmar model. The estimation procedure
by default fits the fixed effect only model, to obtain starting values
for the full model. Random effect variances are given a starting value
of 1, with any covariances given a starting value of 0.

The model estimates a hazard ratio of 0.231 (95\% CI: 0.088, 0.606) for
a female compared to a male of the same age, and a non-statistically
significant age effect. The estimated frailty standard deviation is
0.801 (95\% CI: 0.411, 1.563), indicating a highly heterogenous baseline
hazard function.

We can relax the assumption of proportional hazards by forming an interaction between a covariate of interest and a function of time. \texttt{stmixed} allows an interaction between covariates and a restricted cubic spline function of log time, through the \texttt{tvc()} and \texttt{dftvc()} or \texttt{knotstvc()} options, for example, we form an interaction between log time (\texttt{dftvc(1)}) and \texttt{female}.

\begin{verbatim}
. stmixed age female || patient:, distribution(rp) df(3) tvc(female) dftvc(1)
Random effect M1: Intercept at level patient
variables created: _rcs1_1 to _rcs1_3
variables created for model 1, component 3: _cmp_1_3_1 to _cmp_1_3_1

Fitting fixed effects model:

Fitting full model:

Iteration 0:   log likelihood = -328.64462  
Iteration 1:   log likelihood = -323.89207  (not concave)
Iteration 2:   log likelihood = -323.87879  (not concave)
Iteration 3:   log likelihood = -323.86135  (not concave)
Iteration 4:   log likelihood = -323.82013  (not concave)
Iteration 5:   log likelihood = -323.80934  
Iteration 6:   log likelihood = -323.79736  
Iteration 7:   log likelihood = -323.79734  

Mixed effects survival model                    Number of obs     =         76
Log likelihood = -323.79734
------------------------------------------------------------------------------
             |      Coef.   Std. Err.      z    P>|z|     [95% Conf. Interval]
-------------+----------------------------------------------------------------
_t:          |            
         age |   .0093647   .0112381     0.83   0.405    -.0126615     .031391
      female |  -1.592196    .440563    -3.61   0.000    -2.455683   -.7287083
female#rcs() |   .6843468   .2888028     2.37   0.018     .1183037     1.25039
 M1[patient] |          1          .        .       .            .           .
       _cons |  -.4835611   .5827087    -0.83   0.407    -1.625649    .6585271
-------------+----------------------------------------------------------------
patient:     |            
      sd(M1) |   .5666527    .318761                      .1881428    1.706657
------------------------------------------------------------------------------
    Warning: Baseline spline coefficients not shown - use ml display
\end{verbatim}

Given the statistically significant interaction, we observe evidence of non-proportionality in the effect of \texttt{female}. There is substantial flexibility in being able to model non-proportional hazards in any number of covariates, with differing degrees of freedom.

\subsubsection{Predictions}

\begin{figure}[!h]
    \centering
    \includegraphics[width=0.6\textwidth]{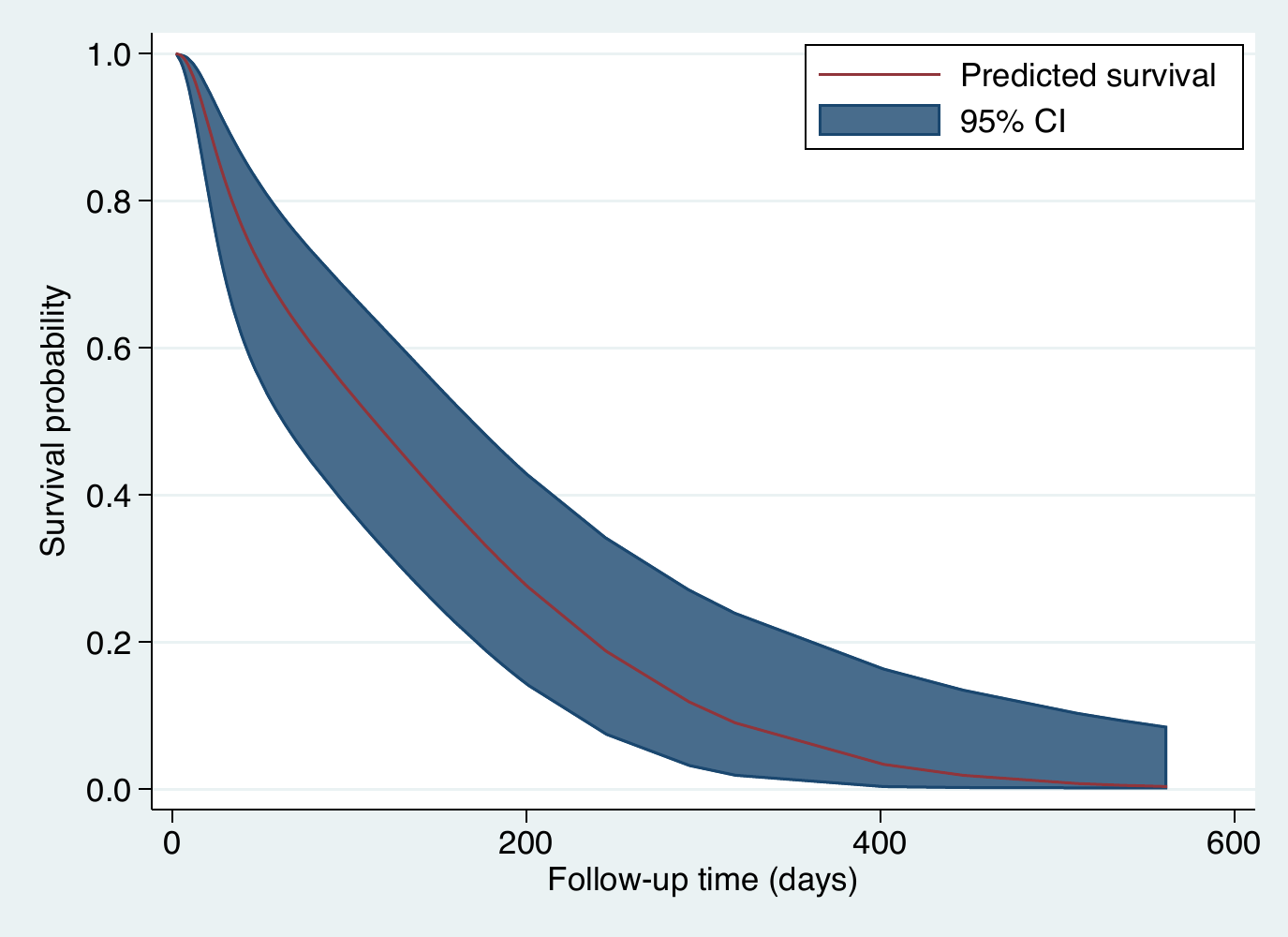}
    \caption{Predicted survival for a female, aged 45, based on the fixed portion of the model.}
    \label{fig1}
\end{figure}

A variety of predictions can be obtained following the fitting of a
model. I can obtain the predicted survival function, shown in Figure
\ref{fig1} with 95\% confidence interval, based on the fixed portion of
the model, for a female aged 45, through use of the \texttt{at()}
option, as follows

\begin{verbatim}
. predict s1, survival ci at(age 45 female 1) 
note: confidence intervals calculated using Z critical values
\end{verbatim}

Which can be plotted by

\begin{verbatim}
. twoway  rarea s1_lci s1_uci _t, sort || line s1 _t, sort      ///
>         ylabel(,angle(h) format(%2.1f))                       ///
>         xtitle("Follow-up time (days)")                       ///
>         ytitle("Survival probability")                        ///
>         legend(order(2 "Predicted survival" 1 "95% CI")       ///
>         ring(0) pos(1) cols(1))
\end{verbatim}

To compare across covariate patterns, for example, to assess the impact
of gender, we can predict restricted mean survival as follows,

\begin{verbatim}
. predict rmst_male  , rmst marginal ci at(age 45 female 0)
note: confidence intervals calculated using Z critical values

. predict rmst_female, rmst marginal ci at(age 45 female 1)
note: confidence intervals calculated using Z critical values
\end{verbatim}

and plotting,

\begin{verbatim}
. twoway  (rarea rmst_male_lci rmst_male_uci _t, sort)                ///
>         (line rmst_male _t, sort)                                   ///
>         (rarea rmst_female_lci rmst_female_uci _t, sort color(%70)) ///
>         (line rmst_female _t, sort)                                 ///
>         , ylabel(,angle(h) format(%2.1f))                           ///
>         xtitle("Follow-up time (days)")                             ///
>         ytitle("Restricted mean survival time")                     ///
>         legend(order(2 "Male" 4 "Female")) 
\end{verbatim}

\begin{figure}[!h]
    \centering
    \includegraphics[width=0.6\textwidth]{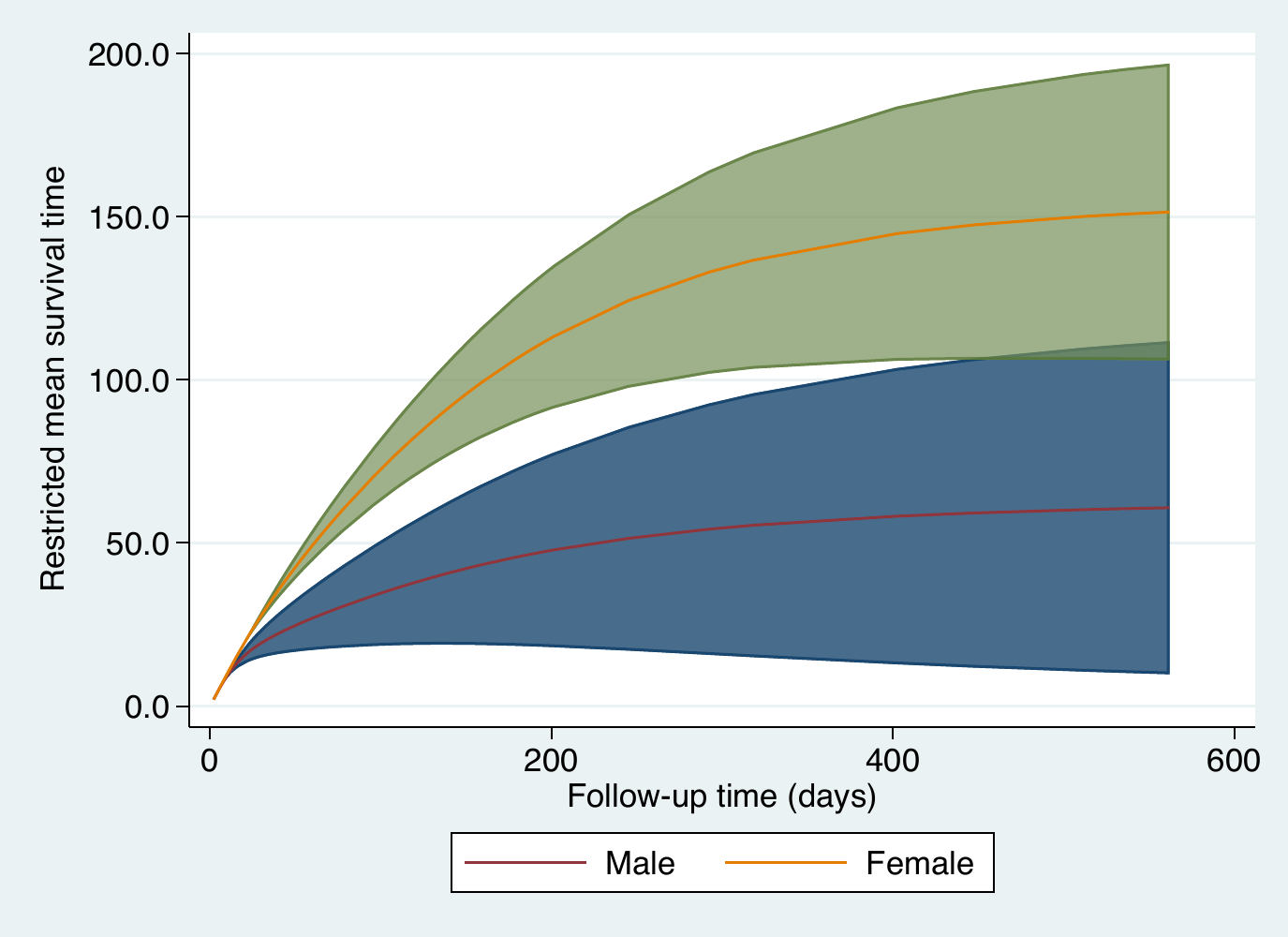}
    \caption{Restricted mean survival for a male or female aged 45.}
    \label{fig2}
\end{figure}

where \ref{fig2} shows restricted mean survival as a function of time,
for a male or female patient with the same age of 45. The impact of
being female is shown clearly, indicating females live substantially
longer than a male of the same age.

\subsection{Individual participant data meta-analysis of survival data}

In this example, I will illustrate a simple approach to simulating
clustered survival data, in the setting of an IPD meta-analysis of
survival data, through the use of the \texttt{survsim} command
\citep{Crowther2012b, CrowtherSurvsim}. I assume a scenario where I have
data from 30 trials, each with 100 patients. Each trial compared a
treatment to a control, with the probability of being assigned to each
arm being 50\%. I assume that the treament effect for each trial comes
from a normal distribution, \(N(-0.5,0.5^2)\), i.e.~an average log
hazard ratio of -0.5 (hazard ratio = \(\exp(-0.5)=0.607\)), with
heterogeneity standard deviation of 0.5. I assume a Weibull baseline
hazard function, with scale and shape parameter values of
\(\lambda=0.1\) and \(\gamma=1.2\), indicating 50.2\% survival in the
control group after 5 years, at which time administrative censoring is
assumed.

\begin{verbatim}
. clear

. // Set seed for reproducibility
. set seed 278945

. // Assume we have 30 trials
. set obs 30
number of observations (_N) was 0, now 30

. // Generate a trial id variable 
. gen trialid = _n

. // Generate trial specifc treatments effect (log hazard ratio) from a 
. // normal distribution with mean 0.5, and std. dev. 0.5
. gen trteffect = rnormal(-0.5,0.5)

. // Assume 100 patients in each trial
. expand 100
(2,970 observations created)

. // Generate 0/1 patient level treatment group indicator
. gen trt = runiform()>0.5

. // Generate a variable containing patient specific treatment 
. // effects for use in simulation
. gen trteffectsim = trt*trteffect

. // Simulate survival times from a Weibull distribution, 
. // incorporating the random treatment effect
. survsim stime died, dist(weibull) lambda(0.1) gamma(1.2)  ///
>                     covariates(trteffectsim 1) maxt(5)

. stset stime, f(died)

     failure event:  died != 0 & died < .
obs. time interval:  (0, stime]
 exit on or before:  failure

------------------------------------------------------------------------------
      3,000  total observations
          0  exclusions
------------------------------------------------------------------------------
      3,000  observations remaining, representing
      1,239  failures in single-record/single-failure data
 11,877.683  total analysis time at risk and under observation
                                                at risk from t =         0
                                     earliest observed entry t =         0
                                          last observed exit t =         5

. stmixed trt || trialid: trt, nocons distribution(weibull)
Random effect M1: trt at level trialid

Fitting fixed effects model:

Fitting full model:

Iteration 0:   log likelihood = -3945.0537  
Iteration 1:   log likelihood = -3943.6472  
Iteration 2:   log likelihood = -3940.1832  
Iteration 3:   log likelihood = -3939.8849  
Iteration 4:   log likelihood = -3939.8835  
Iteration 5:   log likelihood = -3939.8835  

Mixed effects survival model                    Number of obs     =      3,000
Log likelihood = -3939.8835
------------------------------------------------------------------------------
             |      Coef.   Std. Err.      z    P>|z|     [95% Conf. Interval]
-------------+----------------------------------------------------------------
_t:          |            
         trt |  -.4622045   .1240911    -3.72   0.000    -.7054186   -.2189904
trt#M1[tri~] |          1          .        .       .            .           .
       _cons |  -2.415182   .0608592   -39.68   0.000    -2.534464     -2.2959
  log(gamma) |   .2017346   .0264974     7.61   0.000     .1498006    .2536686
-------------+----------------------------------------------------------------
trialid:     |            
      sd(M1) |   .5914661   .0934644                      .4339319    .8061913
------------------------------------------------------------------------------
\end{verbatim}

A key trick to note here is in the \texttt{survsim} command, I included
the variable \texttt{trteffectsim} and assigned it a coefficient value
of 1. This allows you to seamlessly incorporate random effects on
covariates which are included in the linear predictor, multiplied by a
coefficient of 1. Then when the model is fitted using
\texttt{stmixed},the \texttt{trt} variable is used which indicates
treatment group.

In the \texttt{stmixed} model fit, I enter \texttt{trt} as both a fixed
and random effect, but using the \texttt{nocons} option to indicate no
random intercept. This is a rather restrictive model as it assumes that
each trial has the same baseline hazard function. In practice, you may
include the trial id variable in the linear predictor, to allow
proportional trials effects, or indeed stratify by trial membership to
allow separate trials effects \citep{Crowther2012a, Crowther2014a}, or
allow a random intercept at the trial level.

\section{Conclusion}
\label{sec:conc}

In this paper, I have introduced the \texttt{stmixed} command for
multilevel mixed effects survival analysis. \texttt{stmixed} provides
substantial extensions to \texttt{mestreg}, including flexible
spline-based survival models, user-defined survival models, the
extension to relative multilevel survival, simple or complex
time-dependent effects, and \(t\)-distributed random effects. I hope the
wide range of survival models available will be found useful in applied
research. The most up-to-date version of \texttt{stmixed} can be
installed using \texttt{ssc\ install\ stmixed}.

Given that \texttt{stmixed} is essentially a shell file for
\texttt{merlin}, any improvements that are implemented in
\texttt{merlin} will filter up to \texttt{stmixed}.

\section*{About the author}

Michael J. Crowther is an Associate Professor of Biostatistics at the
University of Leicester. He works heavily in methods and software
development, particularly in the field of survival analysis. He is
currently part funded by a MRC New Investigator Research Grant
(MR/P015433/1).

\renewcommand\refname{References}
\bibliography{stmixed_tutorial_arxiv}

\end{document}